\documentclass[11pt,journal,onecolumn]{IEEEtran}

\usepackage{ifpdf}
\usepackage{graphics}
\usepackage{color}
\usepackage{epsfig}
\usepackage{mathptmx}
\usepackage{times}
\usepackage{algpseudocode}
\usepackage{algorithmicx}
\usepackage{amssymb}
\usepackage{amsmath}
\usepackage{amsfonts}
\usepackage{bm}
\usepackage{graphicx}
\usepackage{psfrag}
\usepackage{url}
\usepackage{multirow}
\usepackage{mathrsfs}
\usepackage{enumerate}
\usepackage{midfloat}
\usepackage{pdfpages}
\usepackage[latin1]{inputenc}

\newtheorem{theorem}{Theorem}

\newtheorem{proposition}{Proposition}
\newtheorem{remark}{Remark}

\newcommand{\be}{\begin{equation}}
\newcommand{\ee}{\end{equation}}
\newcommand{\ben}{\begin{equation*}}
\newcommand{\een}{\end{equation*}}
\newcommand{\ba}{\begin{array}}
\newcommand{\ea}{\end{array}}

\newcommand{\defi} { \stackrel{\bigtriangleup}{=} }

\begin{document}

\title{Moving horizon estimation for discrete-time linear systems with binary sensors: algorithms and stability results}

\author{
\IEEEauthorblockN{G. Battistelli$^{a}$, L. Chisci$^{a}$, S. Gherardini$^{a,b}$} \\
\IEEEauthorblockA{$^{a}$Universit\`a  di Firenze, Dipartimento di Ingegneria dell'Informazione (DINFO), Via di Santa Marta 3, 50139 Firenze, Italy. \\
$^{b}$CSDC, Universit\`a  di Firenze, INFN, and LENS, Via G. Sansone 1, I-50019 Sesto Fiorentino, Italy, \\ and QSTAR, Largo E. Fermi 2, I-50125 Firenze, Italy. \\
\{giorgio.battistelli, luigi.chisci, stefano.gherardini\}@unifi.it}
}

\maketitle

\begin{abstract}
The paper addresses state estimation for linear discrete-time systems with binary (threshold) measurements. A \textit{Moving Horizon Estimation} (MHE) approach is followed and different estimators, characterized by two different choices of the cost function to be minimized and/or by the possible inclusion of constraints, are proposed. Specifically, the cost function is either quadratic, when only the information pertaining to the threshold-crossing instants is exploited, or piece-wise quadratic, when all the available binary measurements are taken into account. Stability results are provided for the proposed MHE algorithms in the presence of unknown but bounded disturbances and measurement noise. Performance of the proposed techniques is also assessed by means of simulation examples.
\end{abstract}


\section{Introduction}

Binary (threshold) sensors whose output can take two possible values according to whether the sensed variable exceed or not a given threshold, are nowadays commonly exploited for monitoring/control aims in a wide range of application domains. A non-exhaustive list of existing binary sensors includes: industrial sensors for brushless dc motors, liquid levels, pressure switches; chemical process sensors for vacuum, pressure, gas concentration and power levels; switching sensors for exhaust gas oxygen (EGO or lambda sensors), ABS, shift-by-wire in automotive applications; gas content sensors ($CO$, $CO_2$, $H_2$, etc.) for gas \& oil industry; traffic condition indicators for \textit{asynchronous transmission mode} (ATM) networks; medical sensors/analyses with dichotomous outcomes. In some applications, binary sensors represent the only viable solution for real-time monitoring. In any case, they provide a remarkably more cost-effective alternative to traditional (continuous-valued) sensors at the price of an accuracy deterioration which can, however, be compensated by using many binary sensors (for different variables and/or thresholds) in place of a single one or few traditional sensors. Moreover, binary (threshold) measurements arise naturally in the context of networked state estimation when, in order to save bandwidth and reduce the energy consumption due to data transmission, the measurements collected by each remote sensor are compared locally with a (possibly time-varying) threshold and only information pertaining to the threshold-crossing instants is transmitted to the fusion center. This latter setting falls within the framework of event-based or event-triggered state estimation \cite{BaBeCh,Lazar,likelihood}, and is more challenging as compared to the usually addressed settings due to the minimal information exchange.

The above arguments, as well as the difficulties due to the very limited information provided by binary measurements, has motivated the work on the exploitation of binary measurements for estimation purposes. In particular, \cite{state_reconstruction,Irr-sampling} investigated observability and observer design for linear time-invariant (LTI) continuous-time systems under binary-valued output observations. The work in \cite{Wang1,Wang2} addressed system identification using binary sensors. Specific attention was also devoted to state estimation of hybrid nonlinear systems with binary/quantized sensors \cite{Koutsoukos} and to target tracking with binary sensor networks \cite{Aslam}. A possible solution for coping with the high nonlinearity associated with binary measurements within a stochastic framework is particle filtering \cite{Djuric_2,Ristic}. However such techniques, while effective in many contexts, suffer from the so-called curse of dimensionality (i.e., the exponential growth of the computational complexity as the state dimension increases) and from the lack of guaranteed stability and performance (being based on Monte Carlo integration).

The present paper addresses state estimation for linear discrete-time systems with binary (threshold) output measurements by following a \textit{moving horizon estimation} (MHE) approach. MHE techniques were originally introduced to deal with uncertainties in the system knowledge \cite{Jazwinski} and, in recent years, have gathered an increasing interest thanks to their capability of taking explicitly into account constraints on state and disturbances in the filter design \cite{RaoRawLee01}, and on the possibility of having guaranteed stability and performance even in the nonlinear case \cite{RaRaMa03,NLMHE,AlBaBaZavCDC10}. In fact, MHE has been successfully applied in many different contexts, ranging from switching and large-scale systems \cite{AlBaBaTAC05,GuoHuang13,FaFerrSca10,HabVerh13,SchnHannMarq15} to networked systems \cite{Farina1,Farina2,quantized_measurement}.

In this paper, the state estimation problem with binary measurements is cast in a deterministic framework, in the sense that no probabilistic description of the plant disturbance and noises is supposed to be available. The estimates are computed by minimizing suitable cost functions defined over a given time-horizon (advancing in time) of finite length, possibly subject to linear inequality constraints accounting for the threshold measurements. Specifically, two different approaches are proposed and analyzed. In the first approach, only the threshold-crossing instants are taken into account in the definition of the cost function, by penalizing the distance of the expected continuous outputs (based on the state estimates) from the threshold at those instants. The main advantage of this solution is that the resulting cost function is quadratic. The second approach, instead, exploits all the available information by defining a piece-wise quadratic cost function which accounts for all the available binary measurements, but requires the solution of a convex optimization problem at each time instant. Both unconstrained and constrained MH state estimators will be presented for the two different choices of the cost function and stability results will be proved, assuming unknown but bounded disturbances.

Summarizing, the paper provides the following contributions.
\begin{itemize}
\item Design of novel receding-horizon state estimators for linear discrete-time systems subject to binary (threshold) measurements using either a quadratic or a piecewise quadratic cost function to be minimized and, independently, either including or not constraints.
\item Stability analysis showing that all proposed estimators, irrespectively of the cost being used and of the inclusion of constraints, guarantee an asymptotically bounded estimation error under bounded disturbances and suitable observability assumptions.
\item Performance comparison demonstrating the effectiveness, in terms of both estimation accuracy and computational cost, of our approach.
\end{itemize}
Some of the results of this paper have been preliminarily presented, without proof, in \cite{CDC15}.

The rest of the paper is structured as follows. Section 2 formulates the estimation problem of interest. Section 3 discusses how to solve the problem by means of the MHE approach, with different variants depending on the choice of the cost function as well as on the inclusion or not of constraints. Section 4 deals with the stability analysis of the proposed MH estimators. In section 5, some numerical examples are presented in order to evaluate and compare the proposed estimators. Finally, section 6 ends the paper with concluding remarks and perspectives for future work.

\section{Problem formulation and preliminary considerations}

The following notation will be used throughout the paper: $col(\cdot)$ denotes the matrix obtained by stacking its arguments one on top of the other; $diag(m_{1},\ldots,m_{q})$ denotes the diagonal matrix whose diagonal elements are the scalars $m_{1},\ldots,m_{q}$; further, given a matrix $M$, $vec(M)$ denotes the linear transformation which converts the matrix $M$ into a column vector and
$\|v\|_{M} \defi v' M v$. Finally, $\otimes$ denotes the Kronecker product.

Let us consider the problem of recursively estimating the state of the discrete-time linear dynamical system
\begin{equation}\label{1}
\begin{array}{rcl}
x_{t+1}  & = &  Ax_{t}+Bu_{t}+ w_{t} \\
z_{t}^{i} & = &  C^{i}x_{t}+ v_{t}^{i},\hspace{3mm}i=1,\ldots,p
\end{array}
\end{equation}
from \textit{binary (threshold)} measurements
\begin{equation} \label{3}
\begin{array}{rclcl}
y_{t}^{i} & = & h^i(z_{t}^{i}) & = &  \left\{
\begin{array}{ll} +1, & \mbox{if }  z_{t}^{i} \geq \tau^{i} \\
                         -1,   & \mbox{if }  z_{t}^{i}  < \tau^{i}
\end{array}
\right.
\end{array}
\end{equation}
In (\ref{1})-(\ref{3}): $x_t \in \mathbb{R}^n$ is the state to be estimated; $u_t \in \mathbb{R}^m$ is a known input; $z_t = col \left( z_t^i \right)_{i=1}^p \in \mathbb{R}^p$; $\tau^i$ is the threshold of the $i-$th binary sensor; $A, B, C = col \left( C^i \right)_{i=1}^p$ are matrices of compatible dimensions; $w_t$ and $v_t = col \left( v_t^i \right)_{i=1}^p$ are the process and, respectively, measurement noises assumed \textit{unknown but bounded}. Notice from (\ref{1})-(\ref{3}) that sensor $i$ provides a binary measurement $y_t^i \in \{-1,+1\}$ (two-level measurement quantization) according to whether the noisy linear function of the state $z_t^i = C^i x_t + v_t^i$ falls below or above the threshold $\tau^i$. The problem (\ref{1})-(\ref{3}) clearly includes, as a special instance, the case of quantized sensors with an arbitrary number of levels. In fact, a $d$-level, for generic $d  \geq 2$, quantizer can be easily realized by using $d-1$  binary (threshold) sensors  for the same physical variable but with appropriate different thresholds.
The considered setting with multiple binary sensors (which can measure the same physical variable with different thresholds but also different physical variables) is clearly more general.

It is worth to point out that the system (\ref{1})-(\ref{3}) represents a very special instance of a linear system with output nonlinearity, i.e. a \textit{Wiener} system \cite{Wiener1}. However, due to the discontinuous nature of the measurement function (\ref{3}), all those state estimation techniques for Wiener systems that require a certain smoothness of the output nonlinearity (see for example \cite{Wiener2} and the references therein) cannot be applied. In fact, while general-purpose nonlinear estimators accounting for such a discontinuity (e.g., the particle filter) could be used, the peculiar nature of the considered output nonlinearity deserves special attention and, for optimal exploitation of the poor available information,  the development of ad-hoc receding-horizon estimators that will be presented in the sequel.

Before addressing the estimation problem, some preliminary considerations on the information provided by binary multisensor observations are useful. With this respect, it has been pointed out in \cite{Irr-sampling} that, in the continuous-time case, the information provided by a binary sensor of the form (\ref{3}) is strictly related to the threshold-crossing instants. In fact, in this case, at every instant corresponding to a discontinuity of the binary signal $y^i$, it is known that the signal $z^i$ is equal to the threshold value $\tau^{i}$, implying that the linear measurement $z^i = \tau^i$ is available. Hence, observability with binary sensors for continuous-time linear systems can be analyzed within the more general framework of observability for irregularly sampled systems \cite{Irr-sampling}. In particular, observability can be ensured when the number of threshold-crossing instants (which corresponds to the number of available irregularly sampled linear measurements) is sufficiently large.

The situation is, however, different for discrete-time systems. To see this, consider a generic time instant $k$ in which the binary signal $y_k^i$ changes sign, i.e., $y_{k}^{i}y_{k+1}^{i}<0$. Then, it is not possible to state, as in the continuous-time case, that $z_{k}^{i}$ coincides with the threshold $\tau^i$. Conversely, it can be simply concluded that there exists $\alpha \in [0,1]$ such that
\begin{equation}\label{29}
\alpha \, z_{k}^{i}+(1-\alpha) \, z_{k+1}^{i}=\tau^{i} \, ,
\end{equation}
the exact value of $\alpha$ being clearly unknown and unobservable from the binary measurements. Notice that (\ref{29}) simply states that if the binary output $y_k^i$ switches from discrete time $k$ to $k+1$, then the threshold $\tau^i$ must lie in the interval between $z_k^i$ and $z_{k+1}^i$. In view of (\ref{29}), such discrete time instants $k$ at which the output of some binary sensor changes value will be more appropriately referred to as \textit{output switching} or simply \textit{switching} instants, instead of \textit{threshold-crossing} instants like in the continuous-time case considered in \cite{Irr-sampling}. It is easy to see that (\ref{29}) corresponds to an {\em uncertain} linear measurement
\be\label{4}
\alpha \, z_{k}^{i}+(1-\alpha) \, z_{k+1}^{i} = C^i x_{k} + \delta^i_{k} + \eta^i_{k},
\ee
where $\delta^i_{k} $ is the uncertainty and $\eta^i_{k}$ the measurement noise given by
\begin{equation*}
\begin{split}
&\delta^i_{k} = (1-\alpha)C^{i}(A-I)x_{k}+(1-\alpha)C^{i}Bu_{k}, \\
&\eta^i_{k}   = \alpha\, v_{k}^{i}+(1-\alpha) \, v_{k+1}^{i} + (1-\alpha) \, C^{i} \, w_{k} \, .
\end{split}
\end{equation*}
As a consequence, even in presence of bounded disturbances, the uncertainty associated with the measurement (\ref{29}) depends on $x_k$ and $u_k$.
Recalling that, in general in the context of state estimation for uncertain systems, boundedness of the state trajectories is a prerequisite for the boundedness of the estimation error - see, for instance, the discussion in Section 2.1 of \cite{BlMi} - our attention will be restricted to the case of bounded state and input trajectories by making the following assumption. \vspace{.3cm}

\begin{enumerate}[\bf {A}1]
\item At any time $t$, the vectors $x_t$, $u_t$, $w_t$, $v^i_t, \, i=1, \ldots, p$, belong to the compact sets $X$, $U$, $W$, and $V^i, \, i=1, \ldots,p$, respectively.
\end{enumerate} \vspace{.3cm}

In practice, the compact sets $X$, $U$, $W$, $V^i$ need not be known by the estimator; they will only be used for stability analysis purposes.
\begin{remark}
When the discrete-time system is obtained by sampling a continuous-time one with system matrices ($A_{c},B_{c},C^i$), then the amplitude of the uncertainty $ \delta^i_{k}$ can be related to the sampling interval $T_{s}$. In fact, it turns out that, since in this case $A = e^{A_c T_s}$ and $B=\int_{0}^{T_s} e^{A_c t} \, B \, dt$, $\delta^i_{k}$ vanishes as $T_s$ goes to zero and, in addition, when $T_s$ is small
$ \delta^i_{k}  \approx T_s \left [ (1-\alpha) \, C^{i} \, A_c \, x_{k}+(1-\alpha) \, C^{i} \, B_c \, u_{k} \right ]$.
\end{remark}

\section{Moving horizon estimation for binary sensors}

In order to estimate the state $x_t$ of the linear system $(\ref{1})$ given the binary measurements (\ref{1})-(\ref{3}), a MHE approach is adopted. Then, by considering a sliding window $\mathfrak W_t = \{t-N, t-N+1, \ldots, t\}$, the goal is to find estimates of the state vectors $x_{t-N},\ldots,x_{t}$ on the basis of the information available in  $\mathfrak W_t$ and of the state prediction $\overline{x}_{t-N}$ at the beginning of $\mathfrak W_t$.  Let us denote by $\hat{x}_{t-N|t},\ldots,\hat{x}_{t|t}$ the estimates of $x_{t-N},\ldots,x_{t}$, respectively, to be obtained at any stage $t$.

Following the discussion at the end of the previous section, a first natural approach for constructing a MH estimator would amount to considering the information provided by the switching instants inside the sliding window $\mathfrak W_t$, in order to define the cost-function to be minimized. Accordingly, for any time instant $t \ge N$ and for any sensor index $i$, let us define the set $\mathfrak{I}_{t}^i$ of switching instants as
\be\label{5}
\mathfrak{I}^{i}_t = \{ k \in \mathfrak{W}_t : k+1  \in \mathfrak{W}_t \mbox{ and } y_k^i \, y_{k+1}^i < 0 \} .
\ee
Then, the following least-squares cost function can be defined
\begin{equation}\label{17}
J^A_{t} = \|\hat{x}_{t-N|t}-\overline{x}_{t-N}\|^{2}_{P}+ \sum_{k=t-N}^{t-1}\|\hat{x}_{k+1|t}-A\hat{x}_{k|t}-Bu_{k}\|^{2}_{Q}
+ \sum_{i=1}^{p}\sum_{k\in\mathfrak{I}^{i}_t} \|C^{i} \, \hat{x}_{k|t}-\tau^{i}\|^{2}_{R^{i}},
\end{equation}
where the positive definite matrices $P\in\mathbb{R}^{n\times n}$, $Q\in\mathbb{R}^{n\times n}$ and the positive scalars $R^i , \, i = 1, \ldots p$, are design parameters to be suitably chosen. The first term, weighted by the matrix $P$, penalizes the distance of the state estimate at the beginning of the sliding window from the prediction $\overline{x}_{t-N}$. The second contribution, weighted by the matrix $Q$, takes into account the evolution of the state in terms of the state equation (\ref{1}). Finally, for each sensor $i$ the third term weighted by the scalar $R^i$ penalizes the distances of the expected output (based on the state estimates) $C^{i} \, \hat{x}_{k|t}$ from the threshold $\tau^i$ at the switching instants.
Notice that considering the distance from the threshold at the switching instant is equivalent, for sampled-data systems, to considering the beginning of the time interval $[k T_s, (k+1) T_s]$ in which the threshold crossing happens. As a matter of fact, since for a sampled-data system a binary sensor does not provide a precise information on the threshold crossing instant in
the interval $[k T_s, (k+1) T_s]$, considering the distance from the threshold at the beginning of the time interval is just a choice, not necessarily optimal. As an alternative, with little modifications, one could consider for instance the middle point of the interval. Such modifications would not affect the properties (e.g. stability) of the estimator.

Thus, at each time $t \ge N$, the estimates in the window $\mathfrak W_t$ can be obtained by solving the following optimization problem. \vspace{.3cm}

\textbf{Problem $E_{t}^{A}$:} Given the prediction $\overline{x}_{t-N}$, the input sequence $\{ u_{t-N}, \ldots, u_{t-1} \}$, and the sets $\mathfrak{I}^{i}_t , \, i = 1, \ldots, p$,  find the optimal estimates $\hat{x}^{\circ}_{t-N|t},\ldots,\hat{x}^{\circ}_{t|t}$ that minimize the cost function (\ref{17}). \vspace{.3cm}

Concerning the propagation of the estimation procedure from Problem $E_{t}^{A}$ to Problem $E_{t+1}^{A}$, different prediction strategies may be adopted.
For instance, a first possibility consists of assigning to $\overline{x}_{t-N+1}$ the value of the estimate of $x_{t-N+1}$ made at time instant $t$, i.e., $\bar{x}_{t-N+1} = \hat{x}^{\circ}_{t-N+1|t}$. As an alternative, following \cite{NLMHE}, the state equation of the noise-free system can be applied to the estimate $\hat{x}^{\circ}_{t-N|t}$. In this case, the predictions are recursively obtained by
\begin{equation}\label{18}
\overline{x}_{t-N+1}=A\hat{x}_{t-N|t}^{\circ}+Bu_{t-N},\hspace{3mm}t=N,N+1,\ldots \, .
\end{equation}
Such a recursion is initialized with some a priori prediction $\overline x_0$ of the initial state vector. Hereby, this latter possibility will be adopted as it will facilitate the derivation of the stability results (see Section 4).

The main positive feature of Problem $E_{t}^{A}$ is that it admits a closed-form solution since the cost function (\ref{17}) depends quadratically on the estimates $\hat{x}_{t-N|t},\ldots,\hat{x}_{t|t}$ (for the readers' convenience an explicit expression for the solution is reported in the Appendix). On the other hand, such a cost takes into account only the information pertaining to the switching instants, which, however, is intrinsically uncertain as discussed in the previous section.

In order to overcome such a limitation, a different cost function can be considered by taking into account all the time instants in the sliding window $\mathfrak W_t$. To this end, for any sensor $i = 1 , \ldots, p$, let us define the functions
\begin{equation}
\omega^i(z^{i},y^i)= \left\{ \begin{array}{ll}
1, & \mbox{if} ~ \left( z^i - \tau^i \right) y^i < 0 \\
0, & \mbox{otherwise}
\end{array} \right.
\end{equation}
Suppose now that at time $k$ the sensor $i$ provides a measurement $y^i_k = 1$. Then, the information provided by such a measurement is that the linear measurement $z^i_k$ is above the threshold $\tau^i$, i.e., belongs to the semi-interval $[\tau^i, +\infty)$. Such information can be included in the cost function by means of a term of the form $\omega(C^{i}\hat{x}_{k|t},1)~\|C^{i}\hat{x}_{k|t}-\tau^{i}\|^{2}_{R^{i}}$ which penalizes the distance of the expected output $C^{i}\hat{x}_{k|t}$ from $[\tau^i, +\infty)$. Similarly, in the case $y^i_k = -1$, a term of the form $\omega(C^{i}\hat{x}_{k|t},-1)~\|C^{i}\hat{x}_{k|t}-\tau^{i}\|^{2}_{R^{i}}$ can be used to penalize the distance of the expected output $C^{i}\hat{x}_{k|t}$ from $(-\infty,\tau^i]$. Summing up, the inclusion of such terms gives rise to a cost function of the following form
\begin{equation}\label{36}
J^B_{t} = \|\hat{x}_{t-N|t}-\overline{x}_{t-N}\|^{2}_{P}  + \sum_{k=t-N}^{t-1}\|\hat{x}_{k+1|t}-A\hat{x}_{k|t}-B u_{k}\|^{2}_{Q}
+ \sum_{i=1}^{p}\sum_{k=t-N}^{t} \omega^i (C^{i}\hat{x}_{k|t},y_{k}^{i}) \| C^{i}\hat{x}_{k|t}-\tau^{i}\|^{2}_{R^{i}} \, .
\end{equation}

While a closed-form expression for the global minimum of (\ref{36}) does not exist, since $J_t^B$ is piece-wise quadratic, it is easy to see that the cost $J_t^B$ enjoys some nice properties. In fact, while each function $\omega^i \left( C^i \hat{x}_{k|t}, y_{k}^{i}\right)$ per se is discontinuous, the product $\omega^i \left( C^i \hat{x}_{k|t}, y_{k}^{i}\right)\| C^i \hat{x}_{k|t} - \tau^i \|^2_{R^i}$ is continuous since at the points of discontinuity of
$\omega^i \left( C^i \hat{x}_{k|t}, y_{k}^{i}\right)$, i.e., for $C^i \hat{x}_{k|t} = \tau^i$, the product vanishes. Further, for similar reasons, also the derivative $2 \omega^i \left( C^i \hat{x}_{k|t}, y_{k}^{i}\right) R^i (C^i )' ( C^i \hat{x}_{k|t} - \tau^i )$ of the product turns out to be continuous even at $C^i \hat{x}_{k|t} = \tau^i$.
Thus the product $\omega^i \left( C^i \hat{x}_{k|t}, y_{k}^{i}\right)\| C^i \hat{x}_{k|t} - \tau^i \|^2_{R^i}$ is continuously differentiable on $\mathbb R^n$. Hence, the overall cost function $J_t^B$ is continuously differentiable with respect to the estimates $\hat{x}_{t-N|t},\ldots,\hat{x}_{t|t}$ and also strictly convex (since $P>0$ and $Q>0$). Hence, standard optimization routines can be used in order to find its global minimum. Clearly, since an optimization has to be performed, it is also reasonable to include constraints accounting for the available information on the state trajectory so that the solver can work on a bounded solution set. In particular, in order to preserve convexity, it is advisable to consider a convex set $\mathcal X$ containing $X$ (if $X$ is convex, one can simply set $\mathcal X = X$; in general, choosing $\mathcal X$ as a convex polyhedron is preferable so that only linear constraints come into play). Then, at any stage $t=N,N+1,\ldots$, the following optimization problem has to be solved.  \vspace{.3cm}

\textbf{Problem $E_{t}^{B}$:} Given the prediction $\overline{x}_{t-N}$, the input sequence $\{ u_{t-N}, \ldots, u_{t-1} \}$, the measurement sequences \\
$\{ y^i_{t-N}, \ldots, y^i_t , \, i = 1, \ldots, p \}$, find the optimal estimates $\hat{x}^{\circ}_{t-N|t},\ldots,\hat{x}^{\circ}_{t|t}$ that minimize the cost function (\ref{36}) under the constraints $\hat{x}^{\circ}_{k|t} \in \mathcal X$ for $k= t-N , \ldots,t$.
\vspace{.3cm}

Also in this case, the predictions $\overline{x}_{t-N}$ are supposed to be recursively obtained via equation (\ref{18}) starting from a prior prediction $\overline x_0$. Of course, if no information on the set $X$ is available or if it is preferable to resort to an unconstrained optimization routine, one can simply let $\mathcal X = \mathbb R^n$.

As a final remark, it is worth pointing out that for the two previously presented optimization problems there is a trade-off between estimation accuracy and computational cost. In fact, the cost in Problem $E_{t}^{A}$ is quadratic but accounts only for part of the information provided by the sensors, while Problem $E_{t}^{B}$ accounts for all the available information but requires a convex optimization program to be solved.

Some considerations on the computational complexity of the proposed approaches are in order. The solution of Problem  $E_{t}^{A}$ requires simply the minimization of a strictly convex quadratic form in $(n+1) N$ variables,where $n$ is the plant order. Standard techniques like Gaussian elimination can solve this kind of problems with complexity $O(n^3 N^3)$ but faster algorithms are available. This means that this approach is much computationally cheaper as compared to particle filtering algorithm which usually require in the order of $O(10^n)$ particles to provide satisfactory performance. As for the solution of Problem $E_{t}^{B}$, it entails the minimization of a convex and continuously differentiable piecewise quadratic cost function. It is known that this kind of problems can be solved in finite time by means of sequential quadratic programming \cite{PWCP}. Further, many computationally efficient algorithms are available which are able to handle problems with hundreds of optimization variables \cite{OPT1,OPT2} and enjoys super-linear convergence \cite{bounds}. Nevertheless, application of Problem $E_{t}^{B}$ is possible only when the number $n$ of state variables is not too large and the sampling interval is sufficiently long so as to allow the optimization to terminate. In the other cases, one must resort to Problem $E_{t}^{A}$.

\subsection{Accounting for additional constraints}

Provided that some information on the bounds of the process disturbance $w_t$ and measurement noises $v_t^i $ is available, additional constraints can be considered in the determination of the state estimates. For instance, considering a convex (usually polyhedral) set $\mathcal W$ containing $W$, one can impose the constraints
\begin{equation}\label{con:w}
\hat{x}_{k+1|t}-A\hat{x}_{k|t}-B u_{k} \in \mathcal W \, , \quad k = t-N, \ldots, t-1
\end{equation}
in the solution of the optimization problem.
Moreover, assuming the knowledge of upper bounds $\rho_{V}^{i}$ on the amplitudes $|v_t^i | , \, i=1, \ldots, p$, of the
measurement noises, for each $k$ and each $i$, the constraints
\begin{equation}\label{41a}
\begin{cases}
C^{i}\hat{x}_{k|t}<\tau^{i} +\rho_{V}^{i},\hspace{3mm}\text{if}\hspace{2mm}y_{k}^{i}=-1\\
C^{i}\hat{x}_{k|t}>\tau^{i} - \rho_{V}^{i},\hspace{2mm}\text{if}\hspace{2mm}y_{k}^{i}=1
\end{cases}
\end{equation}
can be imposed. With this respect, it is an easy matter to see that the constraints in (\ref{41a}) define a polyhedron in the state space as summarized in the following proposition (the proof is reported in the Appendix).
\vspace{.3 cm}

\begin{proposition}
Given the vector $\hat{\chi}_{t}= vec \left ( [ \hat x_{t-N|t} \cdots \hat x_{t|t}  ]' \right )$ of the estimates in the observation window,
the constraints in (\ref{41a}), for $k = 0, \ldots, N$ and $i=\, \ldots, p$, can be written in compact form as
\begin{equation}\label{43}
\Gamma_{t}\hat{\chi}_{t}<\gamma_{t},
\end{equation}
where
\begin{equation}\label{11}
\begin{split}
&\Gamma_{t} = \left[\Phi_{t}(C\otimes I_{N})\right]\in\mathbb{R}^{pN\times nN}, \\
&\gamma_{t} = \left[\Phi_{t}vec(\mathcal{T}')+vec(\mathcal{V})\right]\in\mathbb{R}^{pN}, \\
&\Phi_{t}= -diag(y_{t-N}^{1},\ldots,y_{t}^{1},y_{t-N}^{2},\ldots,y_{t}^{2},\ldots,y_{t-N}^{p},\ldots,y_{t}^{p})\in\mathbb{R}^{pN\times pN}, \\
& \mathcal{T}= \begin{bmatrix} \tau^{1} &  \cdots & \tau^{1} \\
\vdots & \vdots & \vdots \\
\tau^{p} & \cdots & \tau^{p}
\end{bmatrix}\in\mathbb{R}^{p\times N}, \; \mathcal{V}= \begin{bmatrix} \rho_{V}^{1} & \cdots & \rho_{V}^{1} \\
\vdots & \vdots & \vdots \\
\rho_{V}^{p} & \cdots & \rho_{V}^{p} \end{bmatrix}\in\mathbb{R}^{p\times N}.
\end{split}
\end{equation}
\end{proposition}\vspace{.3cm}

While the inclusion of the constraints (\ref{con:w}) and (\ref{43}) in the convex optimization problem $E_{t}^{B}$ is natural, in some circumstances it may be interesting to combine them also with the quadratic cost $J_{t}^{A}$. For example, minimizing $J_{t}^{A}$ under the linear constraints (\ref{43}) can be a way to account for the information concerning the non switching instants without the necessity of considering the piece-wise quadratic cost. In fact, this would result in a quadratic programming problem (being the cost quadratic and the constraints linear) for which many efficient solvers are available. It is worth to point out that what is, among the above mentioned options, the best choice clearly depends on the situation under consideration and, in particular, on the available computational resources, on the available information (the bounds $\rho_{V}^{i}$ may be unknown), and on the necessity (or not) of having estimates satisfying the constraints (since clearly this property is guaranteed only if the constraints are taken into account in the estimator design). Nevertheless, in the next Section it will be shown that both costs $J_t^A$ and $J_t^B$ imply some nice stability properties of the resulting MH estimator.

\begin{remark}
Notice that while the considered system dynamics is linear, we do not have access to the linear measurements $z_t = C x_t + v_t$ but rather to the nonlinear (binary) measurements $y_t^i = h^i(z_t^i)$ , for which we cannot apply neither the Kalman filter due to nonlinearity of $h^i(\cdot)$ nor the extended Kalman filter due to the discontinuous nature of $h^i(\cdot)$. It is however worth noting that the simplified quadratic cost $J^A_t$ amounts to considering  a {\em fictitious} linear measurement of the form $C^i x_{k} = \tau^i + \eta^i_k$ for each switching instant $k$ in the observation window. In this case and supposing that no constraints are imposed, the estimates could be computed also via a Kalman-like filter. In all the other cases, i.e. when the piecewise quadratic cost $J^B_t$ is used or constraints are imposed in the optimization, this is no longer possible.
\end{remark}

\section{Stability analysis}

The focus of this section is on the analysis of the stability properties of the state estimators obtained by solving, at each time instant, either Problem $E^A_t$ or $E^B_t$. Specifically, a complete analysis is first provided in the more involved case of Problem $E^B_t$. This will be followed by a short discussion on the main differences in the analysis with respect to Problem $E^A_t$. Notice that the analysis carried out in \cite{NLMHE} for the nonlinear case cannot be directly applied in the present context, since the binary sensors do not satisfy the observability requirement of \cite{NLMHE}. The proofs of all results reported in this section can be found in the Appendix.

For each sensor $i$ and for each time instant $t\geq N$, let us denote by $\Theta_{t}^{i}$ the observability matrix concerning the set $\mathfrak{I}_{t}^{i}$ of the switching instants in the observation window $\mathfrak{W}_t$, i.e,
\begin{equation}
\Theta_{t}^{i}={\rm col}(C^{i}A^{k-t+N})_{k\in\mathfrak{I}_{t}^{i}}.
\label{obs1}
\end{equation}
Then, the observability matrix related to the switchings in $\mathfrak{W}_t$ of all binary sensors is
\begin{equation}
\Theta_{t}={\rm col}(\Theta_{t}^{i})_{i=1}^{p}.
\label{obs2}
\end{equation}
Please notice that the observability matrix defined in (\ref{obs1})-(\ref{obs2}) is actually related to the linear subsystem (\ref{1}), with output $z_t$, of the overall system (\ref{1})-(\ref{3}) considering only those discrete-time instants at which some binary sensor output switches.

The following {\em uniform observability} assumption is needed in order to ensure that enough information is provided by the binary sensors in each window $\mathfrak{W}_t$. \vspace{.3cm}

\begin{enumerate}[\bf {A}1]
\setcounter{enumi}{1}
\item \label{44}
For any $ t\geq N$, ${\rm rank}(\Theta_{t})=n$, with $n=\dim(x_{t})$.
\end{enumerate}\vspace{.3cm}

\begin{remark}
The above uniform observability assumption is made in accordance with the observation that each output switching can be associated with a linear (albeit uncertain) measurement of the form  (\ref{29}).  Hence, each switching instant $k$ can be thought of as a sampling instant for the linear output $z^i_k$.
This means that observability of the system depends crucially on the output switching instants in each observation window which, in turn, clearly depend on the thresholds and of the time window length $N$. In practice, the threshold (or the thresholds when multiple sensors are available)  and the time window length $N$ must be chosen taking into account the system dynamics so as to ensure that such an irregular sampling preserves observability. For instance, when only one binary sensor is available, clearly $N$ should be substantially greater than $2n-1$, with $n=dim(x_t)$, so as to ensure that at least $n$ output switching instants are present in each observation window. While some analytical results on observability under irregular sampling are available \cite{Irr-sampling}, the simplest approach amounts to studying, for instance by numerical simulations, how the observability measure $\delta$ varies as a function of the thresholds and of the time window length $N$. See for instance Figure~\ref{fig:observability_measure} in Section 5 concerning the considered case study. Of course, depending on the system dynamics, time-invariant thresholds may not be sufficient to always ensure uniform observability (think for example to the case of a constant linear output). In these cases, observability can be recovered by making each threshold oscillate in the range of variability of the corresponding continuous output $z^i_t$ with a sufficiently high frequency  and by choosing $N$ so that each observation window contains a sufficient number of threshold oscillation periods. This latter solution is particularly convenient in case $z^i_t$ is a measurement collected by a remote sensor and a time-varying threshold $\tau^i_t$ is used for transmission scheduling. \\ To see this, notice first that sufficient conditions relating the rank of an observability matrix under irregular sampling to the number of samples and to the eigenvalues of the state transition matrix can be found along the lines of \cite{Irr-sampling}. Specifically, for an observable sampled-data system (\ref{1})-(\ref{3}), the observability matrix $\Theta_t$ defined in (\ref{obs1})-(\ref{obs2}) has full rank $n$ if the number of switchings $\nu_t$ in the window $\mathfrak{W}_t$ is such that
\begin{equation}
\nu_t / N \geq 2 (n-1) /N  + \omega_{max} / \pi
\label{cond}
\end{equation}
where $\omega_{max} \defi \max_{\lambda \in sp(A)} \angle \lambda$, $sp(A)$ being the spectrum (set of eigenvalues) of $A$ and $\angle \lambda$
the argument of $\lambda \in \mathbb C$. Notice that $\omega_{max}$ can be interpreted as the bandwidth of the system (in radians).
Hence it turns out that, asymptotically for large $N$, the condition (\ref{cond}) amounts to requiring that the density of switchings $\nu_t / N$ be greater than or equal to $\omega_{max} / \pi < 1$, which represents the system-bandwidth to Nyquist-bandwidth ratio. This means that, for large values of $N$, uniform observability can be ensured even with a single binary sensor by making the threshold oscillations sufficiently fast so as to ensure that the density of switching exceeds $\omega_{max} / \pi < 1$.
\end{remark}

Before stating the main stability results, some preliminary definitions are needed. Given a symmetric matrix $S$, let us denote by $\underline{\lambda}(S)$ and $\overline{\lambda}(S)$ the minimum and maximum eigenvalues of $S$, respectively. Further, given a matrix $M$, let us denote by $\|M\|\triangleq\overline{\lambda}(M'M)^{1/2}$ its norm. Given a generic subset $\Psi$ of an Euclidean space, let us define
$ \rho_{\Psi}  \triangleq\overline  \rm sup_{v \in \Psi } \| v \|$. Given a generic quantity $G^{i}$ related to the $i-$th binary sensor, let us define $\overline{G}\triangleq\max_{i}\|G^{i}\|$ and $\underline{G}\triangleq\min_{i}\|G^{i}\|$. Finally, let us define the {\em uniform observability measure} associated to the matrices $\Theta_t$ as
\[
\delta = \inf_{t \ge N}  \left \| \Theta_t \right\| =  \inf_{t \ge N} {\, \underline{\lambda}(\Theta_{t}'\Theta_{t})^{1/2}} \, .
\]
Notice that, under assumption A2, it can be stated that $\delta >0$. The following result can now be stated. \vspace{.1 cm}

\begin{theorem}\label{theorem_1}
Let assumptions A1 and A2 hold. For each $t \ge N$, let the estimate $\hat x_{t-N,t}^\circ$ be generated by solving Problem $E^B_t$, with $\overline x_{t-N}$ recursively obtained via equation (\ref{18}), and consider the estimation error $e_{t-N} \triangleq x_{t-N} - \hat x_{t-N|t}^\circ$. Then, the weighted norm of the estimation error can be recursively bounded as
\begin{equation}\label{59}
\|e_{t-N}\|^{2}_{P}\leq a_{1}\|e_{t-N-1}\|^{2}_{P}+a_{2},\hspace{3mm}t=N,N+1,\ldots
\end{equation}
where
\begin{equation}\label{60}
\begin{split}
& a_{1}=\frac{b_{1}\|A\|^{2}}{b_{2}} , \\
& a_2 = \frac{c_1 \, \| A-I \|^2 \, \rho^2_{\mathcal X}+ c_2 \, \| B \|^2 \, \rho_U^2 +  c_3 \, \rho_W^2 + c_ 4 \, \overline{\rho}_V^2 }{b_2}, \\
&b_{1}= \frac{\overline{\lambda}(P)}{\underline{\lambda}(P)} \left[4+\frac{d_1}{\underline{\lambda}(Q)}\left(d_2+\overline{R} \right)\right],\hspace{3mm}b_{2}=\left(\frac{1}{2}+\frac{\delta^{2}\underline{R}}{4\overline{\lambda}(P)}\right) \\
\end{split}
\end{equation}
and $c_1$, $c_2$, $c_3$, $c_4$, $d_1$, $d_2$ are suitable constants (given in the proof). In addition, if the weights $Q$ and $R^{i}$, $i=1,\ldots,p$, are selected such that $a_{1}<1$, the norm of the estimation error turns out to be asymptotically bounded in that
\[
\limsup_{t \rightarrow + \infty} \|e_{t-N}\| \le e^{\circ}_{\infty}\triangleq \left ( \frac{a_{2}}{1-a_{1}} \right )^{1/2} \, .
\]
\mbox{   }\hfill $\square$
\end{theorem} \vspace{.3 cm}

The reason for analyzing the estimate at the beginning of the observation window is that, due to the nature of the MHE estimation scheme, the estimate $\hat x^\circ_{t-N|t}$ is used to generate the prediction $\bar x_{t-N+1}$ used at time $t+1$. This makes it possible to recursively write $e_{t-N+1} = x_{t-N+1} - \hat x^\circ_{t-N+1|t+1}$ as a function of $e_{t-N} = x_{t-N} - \hat x^\circ_{t-N|t}$. Notice that even in the noise-free case, i.e., when the process disturbance and the measurement noise are zero and hence $\rho_W = \rho_V = 0$, the asymptotic bound $e_\infty^{\circ}$ on the estimation error does not go to zero due to the presence of the term $c_1 \, \| A-I \|^2 \, \rho^2_{\mathcal X}  + c_2 \, \| B \|^2 \, \rho_U^2$ in $a_2$. Indeed, such a term accounts for the intrinsic uncertainty associated with the threshold-crossing instants in discrete-time as discussed at the end of Section 2 (see equation (\ref{29}) and the subsequent discussion). With this respect, it is worth recalling that, when the discrete-time system under consideration is obtained by sampling a continuous-time system, the quantities $\| A-I \|$ and $\| B \|$ vanish as the sampling interval $T_s$ goes to zero. This means that the smaller is the sampling interval, the smaller turns out to be the asymptotic bound on the estimation error since the information concerning the threshold-crossing instants becomes more precise.

Another important issue concerns the solvability of the stability condition $a_1 <1$. In particular, the following result can be readily proved.
 \vspace{.3cm}
\begin{proposition}\label{proposition2}
Let assumption A2 hold. Then, when $\delta >0$,  it is always possible to select the weights $P$, $Q$ and $R^{i}$, $i=1,\ldots,p$, so that $a_{1}<1$. In particular, for given $Q$ and $R^{i}$, $i=1,\ldots,p$, the condition $a_{1}<1$ can be satisfied by letting $P = \varepsilon \overline{P}$, with $\overline{P}$ any positive definite matrix, provided that $\varepsilon$ is suitably small.
\end{proposition}
\vspace{.3cm}

Hence, if the observability measure $\delta$ is strictly positive, it is sufficient to choose $P$ sufficiently small in order to ensure the satisfaction of the stability condition $a_1 <1$.  This result is in accordance with the well-known results on stability of MHE algorithms which stipulate that stability is ensured provided that the weight on the prediction is sufficiently small \cite{NLMHE}.

\begin{remark}
In the statement of Theorem \ref{theorem_1} the estimates $\hat{x}_{t-N,t}^{\circ}$ are generated by solving Problem $E^{B}_t$, in which the constraints $\hat{x}^{\circ}_{k|t} \in \mathcal X$ for $k= t-N , \ldots,t$ are present. From the practical point of view, including such constraints in the optimization is useful in order to take into account bounds on the state variables in the design of the estimator. In fact,
in many contexts, for instance when the bounds correspond to some physical constraints, providing estimates outside bounds can be meaningless.
On the other hand, in other cases, the bounds on the state variables can be unknown. The proposed approach is flexible enough to deal also with such a case since the inclusion of the constraints is not necessary for the stability of the estimation error dynamics. In fact, in the unconstrained case,
the lower bound of each term $\iota(\hat{x}_{k|t}^{\circ},\hat{x}_{k+1|t}^{\circ})$ (see the Appendix) can be derived as follows
\begin{equation*}
\iota(\hat{x}_{k|t}^{\circ},\hat{x}_{k+1|t}^{\circ}) \geq \|C^{i}\hat{x}_{k|t}^{\circ}-\tau^{i}\|^{2}_{R^{i}}-3(L^{i})^{2}\left(\|A-I\|^{2}\|\hat{x}_{k|t}^{\circ}\|^{2}
+\|B\|^{2}\rho_{U}^{2}+\|w^{\circ}_{k|t}\|^{2}\right)
\end{equation*}
and $\sum_{k\in\mathfrak{I}_{t}^{i}}\|\hat{x}_{k|t}^{\circ}\|^{2}=\|\tilde{l}_{t|t}\|^{2}$, where
$\tilde{l}_{t|t}\triangleq{\rm col}(\hat{x}_{k|t}^{\circ})_{k\in\mathfrak{I}_{t}^{i}}=\Phi_{t}\hat{x}_{t-N|t}^{\circ}+\Gamma_{t}\tilde{u}_{t}+\Lambda_{t}\tilde{w}_{t}^{\circ}$.
It can be readily observed that the matrices $\Phi_{t}$, $\Gamma_{t}$ and $\Lambda_{t}$ are proportional to $\Theta^{i}_{t}$, $H^{i}_{t}$ and $D^{i}_{t}$.
More precisely, $\Theta_{t}^{i}=(C^{i}\otimes I_{n})\Phi_{t}$, $H_{t}^{i}=(C^{i}\otimes I_{n})\Gamma_{t}$ and $D_{t}^{i}=(C^{i}\otimes I_{n})\Lambda_{t}$.
\end{remark}

Consider now the case in which, for each $t \ge N$, the estimate $\hat x_{t-N|t}^\circ$ is generated by solving Problem $E^A_t$, with $\overline x_{t-N}$ recursively obtained via equation (\ref{18}). Notice that, in this case, no constraint is imposed on the estimates $\hat x^\circ_{t-N} , \ldots, \hat x^\circ_{t} $ which can be readily obtained as the unique global minimum of the strictly convex quadratic function $J_t^A$. A close inspection of the proof of Theorem \ref{theorem_1} shows that the same line of reasoning can be applied also for Problem $E^A_t$. The main difference is that, when deriving the lower bound for the optimal cost, each term $\iota(\hat{x}_{k|t}^{\circ},\hat{x}_{k+1|t}^{\circ})$ can be simply replaced with the quantity $\|C^{i}\hat{x}^{\circ}_{k|t}-\tau^{i}\|^{2}$ in accordance with the definition of cost $J^{A}_{t}$. Then an inequality analogous to (\ref{59}) can be derived, with the important difference that, in the definition of the novel $a_2$, $\rho_{\mathcal X}$ can be replaced by $\rho_X$  (which is consistent with the fact that the constraint set $\mathcal X$ is not used in the solution of Problem $E^A_t$).

\begin{remark}
While the foregoing analysis does not account for the possible presence of the additional constraints discussed in Section III-A, analogous results could be easily obtained also when the constraints (\ref{con:w}) and/or (\ref{43}) are imposed in the determination of the state estimates. In this case, the bound on the estimation error turns out to be smaller thanks to the additional information provided by such constraints.
\end{remark}

\begin{remark}
As a final remark, it is pointed out that the extension of the stability results reported here to the case in which the binary measurements are obtained by thresholding nonlinear output maps and/or the system dynamics is nonlinear does not entail particular conceptual difficulties, by combining the analysis of Theorem 1 with that of \cite{NLMHE,AlBaBaZavCDC10}. On the other hand, in this case, establishing a link between the observability properties and the number of threshold crossing instants (see Remark 2) appears more challenging. Further, for nonlinear output maps, the resulting cost functions need not be convex.
\end{remark}

\section{Numerical results}

In this section, numerical examples are presented in order to show the effectiveness of the proposed MHE algorithms for binary measurements.
In particular, two different case-studies will be considered: a first simple example concerning a $2$-mass $2$-spring oscillator and a single binary sensor just for the sake of testing the algorithms' capabilities on a critically observable system, and a second example on a network of $2$-mass $2$-spring oscillators with multiple binary sensors to illustrate a more realistic application of the estimators.

\subsection*{Example 1}

\begin{figure}[h!]
\centering
\includegraphics[scale = 8]{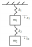}
\caption{$2$-mass $2$-spring mechanical oscillator of example 1.}
\label{fig:oscillators}
\end{figure}

\begin{figure}[h!]
\centering
\includegraphics[scale = 7.5]{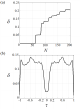}
\caption{Example 1 - (a) Observability measure $\delta$ as a function of the length $N$ of the estimation sliding window (with $\tau = 0.5$). (b) Observability measure $\delta$ as a function of the threshold value $\tau$  (with $N = 100$). The results in (a)-(b) have been evaluated over $100$ Monte Carlo trials.}
\label{fig:observability_measure}
\end{figure}

Let us consider the $2$-mass $2$-spring mechanical system of Fig. \ref{fig:oscillators}.
The state of the system is defined as $x=\left[x_{1},\dot{x}_{1},x_{2},\dot{x}_{2}\right]'$ where $x_{1}$ and $x_{2}$ are the displacements of the two masses from their static equilibrium positions.
Accordingly the system is described by the continuous-time linear state equations $\dot{x}(t) = A_c x(t)$ with
\begin{equation}
\begin{split}
&A_{c}=\begin{bmatrix} 0 & 1 & 0 & 0 \\ -\frac{(k_{1}+k_{2})}{m_{1}} & 0 & \frac{k_{2}}{m_{1}} & 0 \\ 0 & 0 & 0 & 1 \\ \frac{k_{2}}{m_{2}} & 0 & -\frac{k_{2}}{m_{2}} & 0 \end{bmatrix}\end{split}
\end{equation}
where $k_{1}, k_{2}$ are the stiffnesses of the springs and $m_{1}, m_{2}$ the corresponding masses.
The parameters are set to $m_{1}=1=m_{2}=1$ [Kg], $k_{1}=k_{2}=10$ [N/m], and the  continuous-time model is discretized with sampling interval $T_s = 0.1$ [s]. Further, it is assumed that only the displacement $x_2$ (third state component) is measured by a single threshold sensor so that the output matrix turns out to be $C = \left[ 0, 0, 1, 0 \right]$. In all the simulations, the initial state is chosen so as to impose the harmonic motion condition, i.e. $\overline{x}_{0} = [0.618, 0, 1, 0]'$, making the two masses oscillate with the same frequency but different amplitudes within the interval $[-1,1]$; the initial phase of the oscillations is a uniformly distributed random variable. The process disturbance is taken equal to zero, while the measurement noise is a white sequence with uniform distribution in the interval $[-\rho_V,\rho_V]$. In order to tune the proposed MHE algorithms for appropriate performance, the threshold value $\tau$ of the binary sensor and the length $N$ of the estimation sliding window need to be properly selected. To this end, it has been analyzed by means of numerical simulations how the observability measure $\delta$ varies as a function of $N$ and $\tau$, as shown in Fig.~\ref{fig:observability_measure} with a simulation time interval of $50$ [s] and a noise level $\rho_{V} = 0.05$.
As shown in Fig.~\ref{fig:observability_measure}, observability requires  sufficiently large window size ( $N \geq  60$ with $\tau = 0.5$).
Also notice that the observability measure as a function of $N$ has a monotonically increasing behaviour with some characteristic plateaus. Further, it is perfectly symmetric with respect to $\tau$: if the threshold value is outside the range $[-1,1]$ of the system output, then no information is provided by the binary sensor; $\tau = 0$ also implies poor observability as sampling the sinusoid in proximity of zero provides little information about the sinusoid amplitude. From Fig.~\ref{fig:observability_measure}, we chose $N = 100$ and $\tau = 0.5$ for the forthcoming simulation results, so that assumption A2 holds. For the weight matrices we selected $Q = I_{4}$, $R=1$ and $P = \epsilon I_4$ with $\varepsilon<10^{-4}$ in order to satisfy the stability condition $a_1 < 1$ according to Proposition 2.
\begin{figure}[h!]
\centering
\includegraphics[scale = 6]{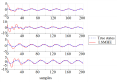}
\caption{Ground truth (dashed blue line) and estimates (solid red line) of the state components versus time, by solving the estimation problem $E_{t}^{A}$.}
\label{fig:sim1}
\end{figure}
Hereafter, for the sake of brevity, the filter obtained by solving at each time instant Problem $E_t^A$ will be referred to as least-squares MHE (LSMHE) algorithm. Similarly, piece-wise MHE (PWMHE) will indicate the filter obtained by solving Problem $E_t^B$. Figs. \ref{fig:sim1} and 4 show the time behavior of the true state variables and of the corresponding estimates in a random simulation with a single binary sensor
by using, respectively, the LSMHE and PWMHE algorithm with measurement noise level $\rho_V = 0.05$ and the same initialization for both algorithms.
\begin{figure}[h!]
\centering
\includegraphics[scale = 6]{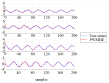}
\caption{Ground truth (dashed blue line) and estimates (solid red line) of the state components versus time, by solving the estimation problem $E_{t}^{B}$.}
\label{fig:sim2}
\end{figure}
As it can be seen, although the estimators are initialized far from the true initial state ($\rho_{\mathcal X} = 5$) and the amount of information exploited in the cost function is limited, the estimates resulting from both algorithms converge to the true trajectories of the systems state vector, and, as  expected, the PWMHE algorithm exhibits much better performance in the transient  thanks to the additional information taken into account in the definition of cost function $J_{t}^{B}$. In order to better appreciate the accuracy of the proposed algorithms and take into account the timescales of the systems, Monte Carlo simulations have been performed by randomly varying the measurement noise realization, the phase of the oscillations for the true state trajectories, and the a priori prediction $\overline{x}_0$, which is randomly generated with uniform distribution in $[-5,5]^4$. For the sake of comparison, also a particle filter with standard sequential importance sampling and $10^3$ particles was tested in the same setting. The results are not reported here because the implemented particle filter was not able to converge and to track the true state.
\begin{figure}[h!]
\centering
\includegraphics[scale = 7]{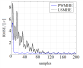}
\caption{Example 1 - Normalized RMSEs of the LSMHE and PWMHE filters, evaluated over $100$ Monte Carlo trials.}
\label{fig:RMSE}
\end{figure}
As performance index, in Fig.~\ref{fig:RMSE} we have adopted a relative error, i.e. the root mean square error (RMSE) normalized by the Euclidean norm of the true system state, where
\begin{equation}\label{64}
\text{RMSE}(t)=\left(\sum_{l=1}^{L}\frac{\|e_{t,l}\|^{2}}{L}\right)^{\frac{1}{2}},
\end{equation}
and $e_{t,l}$ is the state estimation error at time $t$ in the $l-$th simulation run and $L = 100$ is the number of Monte Carlo trials.
Fig.~\ref{fig:RMSE} confirms the effectiveness of the MHE algorithms for state estimation with binary observations.

\begin{figure}[tb]
\centering
\includegraphics[scale = 7.5]{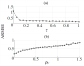}
\caption{(a) ARMSE as a function of the threshold $\tau$ (with $\rho_V = 0.05$ and $N = 100$). (b) ARMSE as a function of the measurement noise level $\rho_V$ (with $\tau = 0.5$ and $N = 100$).
All the results have been evaluated over $100$ Monte Carlo trials.}
\label{fig:ARMSE}
\end{figure}

The computational burden of solving both Problems $E_{t}^{A}$ and $E_{t}^{B}$, as a function of the length $N$ of the estimation sliding window, has been evaluated by means of the CPU time per iteration step (a notebook with an Intel Core i7-2640M CPU @ 2.80 GHz has been used in simulations).
The results are reported in Table~\ref{table_1}.
\begin{table}[tb]
\centering
\begin{tabular}{|c|c|c|}
\hline
N               & LSMHE                   & PWMHE     \\
\hline
1               & 0.50$\cdot10^{-3}$     & 0.25      \\
5               & 0.56$\cdot10^{-3}$     & 0.42      \\
20              & 1.79$\cdot10^{-3}$     & 1.11      \\
35              & 3.23$\cdot10^{-3}$     & 2.07      \\
50              & 5.30$\cdot10^{-3}$     & 3.19      \\
100             & 22.83$\cdot10^{-3}$    & 7.53      \\
150             & 78.90$\cdot10^{-3}$    & 15.70      \\
\hline
\end{tabular}
\caption{CPU time (in [s]) per iteration step for different values of $N$.}
\label{table_1}
\end{table}
Notice that PWMHE is by far more computationally expensive than LSMHE (computing time three orders of magnitude larger in this specific small-size example). As a matter of fact, the solution of Problem $E_{t}^{A}$ can be found analytically by an explicit matrix formula, while for the solution of $E_{t}^{B}$ a convex mathematical programming problem has to be solved. However, it is worth to point out that the PWMHE algorithm has been implemented by using standard functions of the Matlab Optimization Toolbox, without resorting to ad-hoc optimization routines. Hence, we are confident that much faster computing times can be achieved. The dependence of performance on the threshold $\tau$ and the noise level $\rho_V$ is analyzed in Fig.~\ref{fig:ARMSE}, where the ARMSE (i.e., the asymptotic RMSE defined as the average of the RMSE after the transient computed in the time interval $[25,40]$ [s]) is reported for the LSMHE algorithm. Also to compute the ARMSEs in Fig.~\ref{fig:ARMSE}, the RMSEs have been normalized by the Euclidean norm of the true system state.
As observed from Fig.~\ref{fig:ARMSE}a, when the threshold of the sensor is close to zero, performance undergoes a substantial deterioration.
Such a behavior is due to the fact that as $\tau$ goes to zero, the observability measure $\delta$ associated with the output switching instants becomes small. Moreover, as shown in Fig.~\ref{fig:ARMSE}b, the ARMSE decreases almost linearly as the noise level decreases, but, even when the noise is zero, the ARMSE does not go to zero due to the intrinsic uncertainty associated with binary measurements as discussed at the beginning of Section 2.
For similar reasons, even if the process disturbance is taken equal to zero in the simulations, the second term of the cost functions (\ref{17}) and (\ref{36}) does not go to zero, since the estimates need not coincide with the true state
and hence $\hat x_{k+1|t}$ is in general different from $A \hat x_{k|t}$.

\subsection*{Example 2}

\begin{figure}[tb]
\centering
\includegraphics[scale = 5.25]{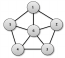}
\caption{Network of six coupled $2$-mass $2$-spring oscillators. Each node of the network has a binary sensor, monitoring the corresponding third state component.}
\label{fig:network}
\end{figure}

\begin{figure}[tb]
\centering
\includegraphics[scale = 7.5]{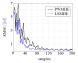}
\caption{Normalized RMSEs of the LSMHE and PWMHE filters, evaluated over $100$ Monte Carlo trials, for a network of six $2$-mass $2$-spring oscillators.}
\label{fig:RMSE_network}
\end{figure}

Finally, in order to numerically assess the performance of the proposed MHE algorithms when the dimensionality of the system state and the number of  binary sensors increase, the network in Fig.~\ref{fig:network} of six coupled $2$-mass $2$-spring oscillators (like the one in Fig. \ref{fig:oscillators}) is considered. It is assumed that each node is equipped with a binary sensor measuring the third component of the local state vector, with threshold belonging to the range $[-1,1]$.
The network dynamics turns out to be described by a discrete-time linear dynamical system with matrices  $A = I_{6}\otimes A_{d} - \gamma  \mathcal{L}\otimes I_{4}$ and $C = I_{6}\otimes [0,0,1,0]$, where: $A_{d} = \exp(A_{c}T_{s})$; $\mathcal{L}$ is the Laplacian matrix of the network; the sampling interval is $T_s = 0.1$ [s]. For the sake of simplicity, we have chosen the same value $\gamma=0.02$ for the coupling constants between all the connected sites, which ensures the synchronization of the system states. Note that synchronization is reached if $\gamma < 0.31685$. The threshold values of the six binary sensors are taken, respectively, equal to $[0.5,0.2,-0.5,-0.8,-0.2,0.3]'$. In all  simulations, the initial state of each $2$-mass $2$-spring system is a uniformly distributed random variable centred around the vector $\overline{x}_{0} = [0.618,0,1,0]'$ with variations of $\pm 5$ for each component, while the measurement noise is a white sequence uniformly distributed in the interval $[-0.05,0.05]$. Moreover, the validity of Proposition~2 for the network is ensured by choosing $\varepsilon = 10^{-5}$ with $\overline{P} = I_{24}$. The duration of each simulation experiment is fixed to $35$ [s], and the corresponding RMSE of the proposed MHE filters is averaged over $100$ Monte Carlo trials. In Fig.~\ref{fig:RMSE_network} the RMSEs, normalized by the Euclidean norm of the true system state, of the LSMHE and PWMHE algorithms are plotted. It can be seen that, also in this case, the PWMHE filter exhibits better performance in the transient, and that the convergence of its estimation error is slower by a factor of approximately $4$ with respect to the case of the single oscillator.

\section{Conclusions}

The paper considers state estimation for linear discrete-time systems, in which the available information is provided by binary multi-sensor observations,
in the presence of unknown but bounded noises affecting both the system and the measurement. Two novel moving-horizon estimators have been introduced, resulting from the minimization of a least-square and a piece-wise quadratic cost function, respectively, with the possible inclusion of constraints. The stability of the estimation error dynamics for the proposed filters has been analyzed and related to the measure of observability
associated with the time instants in which the binary outputs switch. Two simulation examples concerning respectively a single mechanical oscillator and a network of coupled oscillators, have been worked out in order to demonstrate the effectiveness of the proposed approach.

\section*{Acknowledgement}

We acknowledge fruitful discussions with Daniele Mari and Nicola Forti.

\appendix

\section{Proofs}

{\em Closed-form solution of Problem $E_{t}^{A}$:}\\
Let us consider the cost function (\ref{17}). Under the assumption $\bf {A}1$, (\ref{17}) can be written as the following quadratic form:
\begin{equation}\label{A1}
J_{t}^{A} = \hat{Y}_{t-N|t}'M_{t-N}\hat{Y}_{t-N|t}-\hat{Y}_{t-N|t}'D_{t-N}-D_{t-N}'\hat{Y}_{t-N|t}+r_{t-N}= \hat{Y}_{t-N|t}'M_{t-N}\hat{Y}_{t-N|t}+2\hat{Y}_{t-N|t}'U_{t-N}+r_{t-N},
\end{equation}
where $\hat{Y}_{t-N|t}={\rm col}(\hat{x}_{t-N+i|t})_{i=0}^{N}\in\mathbb{R}^{nN}$, $D_{t-N}=-U_{t-N}\in\mathbb{R}^{nN}$ and the matrices $U_{t-N}\in\mathbb{R}^{nN}$, $M_{t-N}\in\mathbb{R}^{nN\times nN}$ are defined as
\begin{equation*}
M_{t-N}=\begin{bmatrix} P+A'QA+\zeta_{j,1} & -A'Q & 0 & \cdots & 0 \\ -QA & Q+A'QA+\zeta_{j,2} & -A'Q & \cdots & 0 \\ \vdots & \vdots & \vdots & \cdots & \vdots \\ 0 & 0 & 0 & \cdots & Q+A'QA+\zeta_{j,N} \end{bmatrix}
\end{equation*}
and
\begin{equation*}
U_{t-N}=\begin{bmatrix} A'QBu_{0}-P\overline{x}_{t-N}-\pi_{j,1} \\ A'Q Bu_{1}-QBu_{0}-\pi_{j,2} \\ \vdots \\ A'Q Bu_{N-1}-QBu_{N-2}-\pi_{j,N-1} \\ A'Q Bu_{N}-QBu_{N-1}-\pi_{j,N}\end{bmatrix},
\end{equation*}
with
\begin{equation*}
\delta_{j,h}^{i}=\begin{cases} 1,\hspace{3mm}\text{if}\hspace{2mm}\exists j\in\mathfrak{I}^{i}_t:j=h \\ 0,\hspace{2mm}\text{else} \end{cases},\hspace{1.5mm}h=1,\ldots,N,
\end{equation*}
and
\begin{equation*}
\begin{split}
&\pi_{j,h}=\sum_{i=1}^{p}\delta_{j,h}^{i}C^{i'}R^{i}\tau^{i},\hspace{3mm}h=1,\ldots,N,\hspace{4mm}
\displaystyle{\zeta_{j,k}=\sum_{i=1}^{p}\delta_{j,h}^{i}C^{i'}R^{i}C^{i}}, \\
&r_{t-N} =\bar{x}_{t-N}'P\bar{x}_{t-N}+\sum_{k=t-N}^{t}u_{k}'B'QBu_{k}+\sum_{i=1}^{p}h_{i}R^{i}\tau^{i^{2}}\in\mathbb{R}, \\
&h_{i} =\dim(\mathfrak{I}^{i}_t).
\end{split}
\end{equation*}
Necessary condition for the minimum of the cost function (\ref{A1}) is
\begin{equation}\label{A2}
\nabla_{\hat{Y}_{t-N|t}}J_{t}^{A}(\hat{Y}_{t-N|t})=2M_{t-N}\hat{Y}_{t-N|t}+2U_{t-N}=0,
\end{equation}
for any $t =N,N+1,\ldots$. Solving (\ref{A2}) as a function of $\hat{x}_{t-N|t}$, we obtain the optimal estimates $\hat{x}^{\circ}_{t-N|t}$, $t=N,N+1,\ldots$ that minimize the cost function (\ref{17}), namely
\begin{equation}\label{A3}
\hat{x}^{\circ}_{t-N|t}=
\begin{bmatrix}I_{n} \underbrace{0 \ldots 0}_{\in\mathbb{R}^{(N-1)n\times n}} \end{bmatrix}M_{t-N}^{-1}D_{t-N},\hspace{3mm}t=N,N+1,\ldots
\end{equation}
Choosing the weighting matrices $P$ and $Q$ as positive semi-definite matrices and $R^{i}>0$, the solution (\ref{A3}) corresponds to a global minimum, since the Hessian matrix $M_{t-N}$ of the cost function is strictly positive definite. As a final remark, notice that there are many equivalent ways of writing the solution of Problem $E_{t}^{A}$ and the particular form presented here is a consequence of the fact
that we consider as optimization variables the state estimates $\hat x_{t-N+i|t}$ for $i = 0, \ldots, N$. An alternative would be to consider as optimization variables the state estimate
$\hat x_{t-N|t}$ at the beginning of the observation interval together with the estimates of the process disturbance $\hat w_{t-N+i|t} = \hat x_{t-N+i+1|t}  - A \hat x_{t-N+i|t} - B u_{t-N+i} $ for
$i = 0, \ldots, N-1$. In this case, each $\hat x_{t-N+i|t}$  would be written as a function of $\hat x_{t-N|t}$ and the observability matrix would explicitly appear in the solution.
\mbox{ } \hfill $\square$
\vspace{.5cm}

{\em Proof of Proposition 1:}\\
For each $k=t-N,\ldots,t-1$, we initially introduce the constraints for the $i-$th measurement equation, $i=1,\ldots,p$:
\begin{equation}\label{B1}
\begin{cases}
C^{i}\hat{x}_{k|t}<\tau^{i}+\rho_{V}^{i},\hspace{3mm}\text{if}\hspace{3mm}y_{k}^{i}=-1 \\
C^{i}\hat{x}_{k|t}>\tau^{i}-\rho_{V}^{i},\hspace{3mm}\text{if}\hspace{3mm}y_{k}^{i}=1
\end{cases}
\end{equation}
The system (\ref{B1}) is equivalent to the inequality
\begin{equation}\label{B2}
y_{k}^{i}(C^{i}\hat{x}_{k|t}+y_{k}^{i}\rho_{V}^{i})>y_{k}^{i}\tau^{i}.
\end{equation}
Observing that $(y_{k}^i)^{2}=1$, $\forall k=t-N,\ldots,t-1$, we obtain
\begin{equation*}\label{B3}
y_{k}^{i}(C^{i}\hat{x}_{k|t}-\tau^{i})+\rho_{V}^{i}>0,\hspace{3mm}k=t-N,\ldots,t-1.
\end{equation*}
If we define $\phi_{k}=diag(y_{k}^{i})\in\mathbb{R}^{p\times p}$, $i=1,\ldots,p$, $\tau_{p}={\rm col}(\tau^{i})_{i=1}^{p}\in\mathbb{R}^{p}$ and $\nu={\rm col}(\rho_{V}^{i})_{i=1}^{p}\in\mathbb{R}^{p}$, then we can write
\begin{equation*}
\phi_{k}(C\hat{x}_{k|t}-\tau_{p})+\nu>0,
\end{equation*}
since $\phi_{k}'\phi_{k}=I_{p}$. Moreover, introducing the matrices $\Phi_{t}$, $\mathcal{T}$ and $\mathcal{V}$ as in (\ref{11}), the constraints (\ref{B1}) can be written in matrix form, namely
\begin{equation}\label{B4}
\Phi_{t}~vec\left[(C\hat{X}_{t}-\mathcal{T})'\right]<vec(\mathcal{V})
\end{equation}
where $\hat{X}_{t}=\left[\hat{x}_{t-N|t},\ldots,\hat{x}_{t|t}\right]'$. Observing that $vec\left[(C\hat{X}_{t})'\right]\equiv(C\otimes I_{n})vec\left(\hat{X}_{t}'\right)$, it can be noted that  (\ref{B4}) is equal to  (\ref{43}),
so that the proposition is proved. \hfill $\square$
\vspace{.5cm}

{\em Proof of Theorem 1:}\\
Some preliminary definitions are needed. Notice first that, while the function $\omega(C^{i}x,y)\|C^{i}x-\tau^{i}\|$ is not differentiable for $C^{i}x=\tau^{i}$, for $C^{i}x \ne \tau^{i}$ one has
\[
\frac{\partial}{\partial x}  \omega(C^{i}x,y)\|C^{i}x-\tau^{i}\| = \left \{ \begin{array}{ll} 0, & \mbox{ if } y(C^{i}x-\tau^{i}) > 0 \, , \\ - y \, C^i , & \mbox{ if } y(C^{i}x-\tau^{i}) < 0 \, .\end{array} \right\} .
\]
Hence $\omega(C^{i}x,y)\|C^{i}x-\tau^{i}\|$ is globally Lipschitz with Lipschitz constant $L^{i} = \| C^i \|$, for $i=1,\ldots,p$. Further, consider for each sensor $i$ and each sliding window $\mathfrak W_t$, the vector $\tilde{z}^{i}_{t|t}=col(C^{i}\hat{x}_{k|t}^{\circ})_{k\in\mathfrak{I}_{t}^{i}}$. Then, we can write
\begin{equation*}\label{49}
\tilde{z}^{i}_{t|t}=\Theta_{t}^{i}\hat{x}_{t-N|t}^{\circ}+H_{t}^{i}\tilde{u}_{t}+D_{t}^{i}\tilde{w}_{t}^{\circ},
\end{equation*}
where
\begin{equation*}\label{48}
\begin{split}
&\tilde{u}_{t}=col(u_{k})_{k\in[t-N,t]}, \\
&w_{k|t}^{\circ}=\hat{x}_{k+1|t}^{\circ}-A\hat{x}_{k|t}^{\circ}-Bu_{k}, \\
&\tilde{w}_{t}^{\circ}=col(w_{k|t}^{\circ})_{k\in[t-N,t]},
\end{split}
\end{equation*}
and $H_{t}^{i}$ and $D_{t}^{i}$ are suitable matrices. Let $\phi^i$ be defined as
$\sup_{t \ge N} \overline{\lambda}({D}_{t}^{i\hspace{0.5mm}'} {D}^i_{t})^{1/2}$. Clearly, $\phi^i$ is finite since  $D^i_t$ can assume only a finite number of configurations in the estimation window.

Let us now consider the estimation error as $e_{t-N}= x_{t-N}-\hat{x}_{t-N}^{\circ}$; the aim is to find a lower and an upper bound for the optimal cost
\begin{equation}\label{45c}
J_{t}^{\circ}
=\|\hat{x}_{t-N|t}^{\circ}-\overline{x}_{t-N}\|^{2}_{P}+\sum_{k=t-N}^{t-1}\|\hat{x}_{k+1|t}^{\circ}-A\hat{x}_{k|t}^{\circ}-Bu_{k}\|^{2}_{Q}
+\sum_{i=1}^{p}\sum_{k=t-N}^{t}\omega(z_{k}^{i},y_{k}^{i})\|C^{i}\hat{x}_{k|t}^{\circ}-\tau^{i}\|^{2}_{R^{i}}.
\end{equation}
to derive a bounding sequence on the norm of the estimation error. \\ \\
\textit{-- Upper bound on the optimal cost $J_{t}^{\circ}$:} \\ \\
For the optimality of the cost function $J^{\circ}_{t}$, we have $\left.J^{\circ}_{t}\leq J_{t}^{B}\right|_{\hat{x}_{k|t}=x_{k},\hspace{1mm}k\in\mathfrak W_{t}}$ and hence
\begin{equation}\label{46}
J_{t}^{\circ} \leqslant \|x_{t-N}-\overline{x}_{t-N}\|^{2}_{P}
+\sum_{k=t-N}^{t-1}\|w_{k}\|^{2}_{Q}+\sum_{i=1}^{p}\sum_{k=t-N}^{t}\omega(z_{k}^{i},y_{k}^{i})\|z_{k}^{i}-\tau^{i}\|^{2}_{R^{i}}.
\end{equation}
The discontinuous function $\omega(z_{k}^{i},y_{k}^{i})$ is non zero if and only if $z_{k}^{i}-\tau^{i} \in V^{i}$, i.e. if the system output is close to the $i-$th sensor threshold and the measurement noise makes the sensor detection incoherent with the system evolution. \\
Thus, the upper bound (\ref{46}) can be rewritten as
\begin{equation}\label{47}
J_{t}^{\circ}\leq\|x_{t-N}-\overline{x}_{t-N}\|^{2}_{P}+N\overline{\lambda}(Q)\rho^{2}_{W}+p(N+1)\overline{R}\overline{\rho}^{2}_{V}.
\end{equation}\\
\textit{-- Lower bound on the optimal cost $J_{t}^{\circ}$:} \\ \\
Let us consider a time instant $k\in\mathfrak{I}_{t}^{i}$ and suppose, for the sake of notational simplicity, that $y^{i}_{k}=1$ and $y^{i}_{k+1}=-1$
(up-down threshold crossing). Note that the dual case can be analysed in a similar way.
Thus, in the cost function $J_{t}^{\circ}$ the following contribution is present:
\begin{eqnarray*}\label{50}
\iota(\hat{x}_{k|t}^{\circ},\hat{x}_{k+1|t}^{\circ})&\triangleq&\omega(z_{k}^{i},1)\|C^{i}\hat{x}_{k|t}^{\circ}-\tau^{i}\|^{2}_{R^{i}} +\omega(z_{k+1}^{i},-1)\|C^{i}\hat{x}_{k+1|t}^{\circ}-\tau^{i}\|^{2}_{R^{i}}\nonumber \\
&=&\underbrace{\left[\omega(z_{k}^{i},1)+\omega(z_{k}^{i},-1)\right]}_{=1,\text{ by definition}}\|C^{i}\hat{x}_{k|t}^{\circ}-\tau^{i}\|^{2}_{R^{i}}
+\omega(z_{k+1}^{i},-1)\|C^{i}\hat{x}_{k+1|t}^{\circ}-\tau^{i}\|^{2}_{R^{i}} - \omega(z_{k}^{i},-1)\|C^{i}\hat{x}_{k|t}^{\circ}-\tau^{i}\|^{2}_{R^{i}},
\end{eqnarray*}
where
\begin{equation*}\label{69}
\omega(z_{k+1}^{i},-1)\|C^{i}\hat{x}_{k+1|t}^{\circ}-\tau^{i}\|^{2}_{R^{i}}-\omega(z_{k}^{i},-1)\|C^{i}\hat{x}_{k|t}^{\circ}-\tau^{i}\|^{2}_{R^{i}}
\leq(L^{i})^{2}\|(A-I)\hat{x}_{k|t}^{\circ}+Bu_{k}+w^{\circ}_{k|t}\|^{2}.
\end{equation*}
Since $\hat{x}^{\circ}_{k|t}\in\mathcal X$ for $k=t-N,\ldots,t$, it can be stated that each term $\iota(\hat{x}_{k|t}^{\circ},\hat{x}_{k+1|t}^{\circ})$ has a lower bound, such that
\begin{equation*}\label{51}
\iota(\hat{x}_{k|t}^{\circ},\hat{x}_{k+1|t}^{\circ})\geq\|C^{i}\hat{x}_{k|t}^{\circ}-\tau^{i}\|^{2}_{R^{i}}-3(L^{i})^{2}\left(\|A-I\|^{2}\rho_{\mathcal X}^{2}
+\|B\|^{2}\rho_{U}^{2}+\|w^{\circ}_{k|t}\|^{2}\right).
\end{equation*}
Since $k$ is switching instant, hence
\begin{equation*}
y^{i}_{k}=h^{i}(C^{i}x_{k}+v^{i}_{k})=1
\end{equation*}
and
\begin{equation*}
y^{i}_{k+1}=h^{i}(C^{i}Ax_{k}+C^{i}Bu_{k}+C^{i}w_{k}+v^{i}_{k+1})=-1,
\end{equation*}
i.e. there exists $\alpha\in[0,1]$ such that
$\alpha z^{i}_{k}+(1-\alpha)z^{i}_{k+1}=\tau^{i}$, from which
\begin{equation*}\label{52}
\tau^{i}=C^{i}x_{k}+\zeta^{i}_{k},
\end{equation*}
where $\zeta^{i}_{k}=\delta^{i}_{k}+\eta^{i}_{k}$. Then,
\begin{equation*}\label{54}
\|C^{i}\hat{x}_{k|t}^{\circ}-\tau^{i}\|^{2}_{R^{i}} = \|C^{i}\hat{x}_{k|t}^{\circ}-C^{i}x_{k}-\zeta^{i}_{k}\|^{2}_{R^{i}}
\geq \frac{1}{2}\|C^{i}\hat{x}_{k|t}^{\circ}-C^{i}x_{k}\|^{2}_{R^{i}}-\|\zeta^{i}_{k}\|^{2}_{R^{i}},
\end{equation*}
where
\begin{equation*}\label{55}
\|\zeta^{i}_{k}\|^{2}_{R^{i}} \leq 4R^{i}\left(\|C^{i}\|^{2}\|A-I\|^{2}\rho_{\mathcal X}^{2}+\|C^{i}\|^{2}\|B\|^{2}\rho_{U}^{2}
+ \|C^{i}\|^{2}\rho_{W}^{2}+(\rho_{V}^{i})^{2}\right).
\end{equation*}
Summarizing the previous results, if we consider $\forall i$ only the instants $k\in\mathfrak{I}_{t}^{i}$, we obtain
\begin{equation*}\label{56}
J_{t}^{\circ}\geq\|\hat{x}_{t-N|t}^{\circ}-\overline{x}_{t-N}\|^{2}_{P}+\sum_{i=1}^{p}\sum_{k\in\mathfrak{I}_{t}^{i}}
\left(\|C^{i}\hat{x}_{k|t}^{\circ}-C^{i}x_{k}\|^{2}_{R^{i}}\right)-\beta_{t}-\sigma_{t},
\end{equation*}
where
\begin{equation*}
\beta_{t} = \sum_{i=1}^{p}\sum_{k\in\mathfrak{I}_{t}^{i}}\left[4R^{i}\left(\|C^{j}\|^{2}\|A-I\|^{2}\rho_{\mathcal X}^{2}+\|C^{i}\|^{2}\|B\|^{2}\rho_{U}^{2}
+ \|C^{i}\|^{2}\rho_{W}^{2}+(\rho_{V}^{i})^{2}\right)+3(L^{i})^{2}(\|A-I\|^{2}\rho_{\mathcal X}^{2}+
\|B\|^{2}\rho_{U}^{2})\right]
\end{equation*}
and $\displaystyle{\sigma_{t}=\sum_{i=1}^{p}\sum_{k\in\mathfrak{I}_{t}^{i}}3(L^{i})^{2}\|w_{k|t}^{\circ}\|^{2}}$ are quantities with an upper bound. Indeed, it can be stated that:
\begin{equation*}
\beta_{t} \leq 4p(N+1)\overline{R}\left(\overline{C}^{2}\|A-I\|^{2}\rho_{\mathcal X}^{2}+\overline{C}^{2}\|B\|^{2}\rho_{U}^{2}+\overline{C}^{2}\rho_{W}^{2}+\overline{\rho}_{V}^{2}\right)
+ 3p(N+1)\overline{L}^{2}\left(\|A-I\|^{2}\rho_{\mathcal X}^{2}+\|B\|^{2}\rho_{U}^{2}\right)=\breve{\beta}_{t}
\end{equation*}
and
\begin{equation*}
\sigma_{t}\leq3p\left(\max_{i}L^{i}\right)^{2}\sum_{k=t-N}^{t-1}\|w_{k|t}^{\circ}\|^{2}
\leq\frac{3p\overline{L}^{2}}{\underline{\lambda}(Q)}\left[\|\hat{x}_{t-N|t}-
\overline{x}_{t-N}\|^{2}_{P}+N\overline{\lambda}(Q)\rho_{W}^{2}+p(N+1)\overline{R}\overline{\rho}_{V}^{2}\right]=\breve{\sigma}_{t}.
\end{equation*}
To conclude the calculation of the lower bound, let us define
$\tilde{z}^{i}_{t}\triangleq col(z_{k})_{k\in\mathfrak{I}_{t}^{i}}$ and $\tilde{R}^{i}\triangleq R^{i}I_{|\mathfrak{I}_{t}^{i}|}$ and write
\begin{equation*}
\psi_{t}\triangleq\sum_{i=1}^{p}\sum_{k\in\mathfrak{I}_{t}^{i}}\left(\|C^{i}\hat{x}_{k|t}^{\circ}-C^{i}x_{k}\|^{2}_{R^{i}}\right)=\sum_{i=1}^{p}\|\tilde{z}^{i}_{t|t}-\tilde{z}^{i}_{t}\|^{2}_{\tilde{R}^{i}}= =\sum_{i=1}^{p}\|\Theta_{t}^{i}\hat{x}_{t-N|t}^{\circ}+H^{i}_{t}\tilde{u}_{t}+D_{t}^{i}\tilde{w}_{t}^{\circ}-\Theta_{t}^{i}x_{t-N}-
H^{i}_{t}\tilde{u}_{t}-D^{i}_{t}\tilde{w}_{t}-\tilde{v}^{i}_{t}\|^{2}_{\tilde{R}^{i}},
\end{equation*}
with $\tilde{w}_{t}\triangleq col(w_{k})_{k\in[t-N,t]}$ and $\tilde{v}^{i}_{t}\triangleq col(v_{k}^{i})_{k\in\mathfrak{I}_{t}^{i}}$. Hence,
\begin{equation*}
\psi_{t} \geq \sum_{i=1}^{p}\left(\frac{1}{4}\|\Theta_{t}^{i}(\hat{x}_{t-N|t}^{\circ}-x_{t-N})\|^{2}_{\tilde{R}^{i}}-\|D_{t}^{i}\tilde{w}_{t}^{\circ}\|^{2}_{\tilde{R}^{i}}- \|D_{t}^{i}\tilde{w}_{t}\|^{2}_{\tilde{R}^{i}}
- \|\tilde{v}^{i}_{t}\|^{2}_{\tilde{R}^{i}}\right)\geq\frac{1}{4}\|\Theta_{t}(\hat{x}_{t-N|t}^{\circ}-x_{t-N})\|^{2}_{\tilde{R}}-\breve{\mu}_{t},
\end{equation*}
where
\begin{eqnarray*}
\mu_{t}&=&\sum_{i=1}^{p}R^{i}\left[\|D_{t}^{i}\|^{2}\left(\|\tilde{w}_{t}^{\circ}\|^{2}+\rho_{W}^{2}\right)+(\rho_{V}^{i})^{2}\right]\nonumber \\
&\leq&p\overline{R}\left[\frac{\overline{\phi}^{2}}{\underline{\lambda}(Q)}\left(\|x_{t-N}-\overline{x}_{t-N}\|^{2}_{P}+N\overline{\lambda}
(Q)\rho_{W}^{2}+p(N+1)\overline{R}\overline{\rho}_{V}^{2}\right)
+\overline{\phi}^{2}\rho_{W}^{2}+\overline{\rho}_{V}^{2}\right]=\breve{\mu}_{t},
\end{eqnarray*}
i.e.
\begin{equation*}
\psi_{t}\geq\frac{\delta^{2}\underline{R}}{4\overline{\lambda}(P)}\|\hat{x}_{t-N|t}^{\circ}-x_{t-N}\|^{2}_{P}-\breve{\mu}_{t}=
\frac{\delta^{2}\underline{R}}{4\overline{\lambda}(P)}\|e_{t-N}\|_{P}^{2}-\breve{\mu}_{t}.
\end{equation*}
In conclusion
\begin{equation}\label{57}
J_{t}^{\circ}\geq\|\hat{x}_{t-N|t}^{\circ}-\overline{x}_{t-N}\|^{2}_{P}+\frac{\delta^{2}\underline{R}}{4\overline{\lambda}(P)}\|e_{t-N}\|_{P}^{2}
-\breve{\beta}_{t}-\breve{\sigma}_{t}-\breve{\mu}_{t}.
\end{equation}
Now we can exploit the bounds on the optimal cost $J_{t}^{\circ}$ in order to obtain a bounding sequence on the norm of the estimation error. More specifically, combining (\ref{47}) and (\ref{57}), we derive the following inequality:
\begin{equation}\label{58}
\|\hat{x}_{t-N|t}^{\circ}-\overline{x}_{t-N}\|^{2}_{P}+\frac{\delta^{2}\underline{R}}{4\overline{\lambda}(P)}\|e_{t-N}\|_{P}^{2}
\leq\breve{\beta}_{t}+\breve{\sigma}_{t}+\breve{\mu}_{t}+\|x_{t-N}-\overline{x}_{t-N}\|^{2}_{P}+N\overline{\lambda}
(Q)\rho_{W}^{2}+p(N+1)\overline{R}\overline{\rho}_{V}^{2}.
\end{equation}
But, noting that
\begin{equation*}
\|\hat{x}_{t-N|t}^{\circ}-\overline{x}_{t-N}\|^{2}_{P}\geq\frac{1}{2}\|e_{t-N}\|_{P}^{2}-\|x_{t-N}-\overline{x}_{t-N}\|^{2}_{P}
\end{equation*}
and
\begin{equation*}
x_{t-N}-\overline{x}_{t-N} =Ae_{t-N-1}+w_{t-N-1},
\end{equation*}
namely
\begin{equation*}
\|x_{t-N}-\overline{x}_{t-N}\|^{2}_{P}\leq 2\left(\|A\|^{2}_{P}\|e_{t-N-1}\|^{2}_{P}+\overline{\lambda}(P)\rho_{W}^{2}\right),
\end{equation*}
inequality (\ref{58}) can be rewritten as
\begin{equation*}
\|e_{t-N}\|^{2}_{P}\leq a_{1}\|e_{t-N-1}\|^{2}_{P}+a_{2},
\end{equation*}
where the coefficients $a_{1}$ and $a_{2}$ are defined as in (\ref{60}) and
\begin{equation*}
\begin{split}
&d_{1}=2p\overline{\phi}^{2},\hspace{4mm}d_{2}=3\overline{L}^{2}\bar{\phi}^{-2}, \\
&c_{1}=c_{2}=p(N+1)\left(4\overline{R}~\overline{C}^{2}+3\overline{L}^{2}\right), \\
&c_{3}=b_{1}+N\overline{\lambda}(Q)\left(\frac{b_{1}}{2\overline{\lambda}(P)}-1\right)+p\overline{R}\left[4(N+1)\overline{C}^{2}+\overline{\phi}^{2}\right], \\
&c_{4}=p(N+1)\overline{R}\left(\frac{b_{1}}{2\overline{\lambda}(P)}-1\right)+p\overline{R}(4N+5).
\end{split}
\end{equation*}
Since $a_{2}$ is a positive scalar, if we further impose that $a_{1}<1$, the asymptotic upper bound $e^{\circ}_{\infty}$ can be easily derived, in that
\begin{equation*}\label{62}
\|e_{t}\|^{2}_{P}<a_{1}^{t}\|e_{0}\|^{2}_{P}+a_{2}\sum^{t-1}_{j=0}a_{1}^{j},
\end{equation*}
which tends to $a_{2}/(1-a_{1})$ as $t\rightarrow\infty$. \hfill \vspace{.1cm} $\square$
\vspace{.5cm}

{\em Proof of Proposition 2:}
Notice first that the stability condition $a_1 < 1$ can be rewritten as
\[
\frac{\overline{\lambda}(P)}{\underline{\lambda}(P)} \left[4+\frac{d_1}{\underline{\lambda}(Q)}\left(d_2+\overline{R} \right)\right] \| A \|^2
\le \left(\frac{1}{2}+\frac{\delta^{2}\underline{R}}{4\overline{\lambda}(P)}\right) \, .
\]
By letting $P = \varepsilon \overline{P}$, with $\overline{P}$ any positive definite matrix, the above inequality becomes
\[
\frac{\overline{\lambda}(\overline{P})}{\underline{\lambda}(\overline{P})} \left[4+\frac{d_1}{\underline{\lambda}(Q)}\left(d_2+\overline{R} \right)\right] \| A \|^2
\le \left(\frac{1}{2}+\frac{\delta^{2}\underline{R}}{4\, \varepsilon \, \overline{\lambda}(\overline{P})}\right) \, .
\]
It can be seen that the left-hand side of such an inequality does not depend on $\varepsilon$, whereas the right-hand side goes to infinity as $\varepsilon$ goes to $0$, provided that $\delta^2 >0$.
Hence, when $\delta^2 >0$, it is always possible to ensure that the stability condition $a_1 < 1$ holds by taking any $Q$, $R_i$, $i = 1, \ldots,p$, $\overline{P}$, and then choosing $\varepsilon$
suitably small.
\mbox{ } \hfill $\square$


\vspace{.1cm}

\begin{thebibliography}{10}

\bibitem{BaBeCh}
Battistelli, G. and Benavoli, A. and Chisci, L.
\newblock Data-driven communication for state estimation with sensor networks.
\newblock {\em Automatica}, vol. 48, pp. 926--935, 2012.

\bibitem{Lazar}
Sijs, J. and Lazar, M.
\newblock Event-based state estimation with time synchronous updates.
\newblock {\em IEEE Trans. on Automatic Control}, vol. 57, pp. 2650--2655, 2012.

\bibitem{likelihood}
Shi, D. and Chen, T. and Shi, L.
\newblock Event-triggered maximum likelihood state estimation.
\newblock {\em Automatica}, vol. 50, pp. 247--254, 2014.

\bibitem{state_reconstruction}
Wang, L.Y. and Xu, G. and Yin, G.G.
\newblock State reconstruction for linear time-invariant systems with binary-valued output observations.
\newblock {\em Systems and Control Letters}, vol. 57, pp. 958--963, 2008.

\bibitem{Irr-sampling}
Wang, L.Y. and Li, G.G. and Guo, L. and Xu, C.-Z.
\newblock State observability and observers of linear-time-invariant systems under irregular sampling and sensor limitations.
\newblock {\em IEEE Trans. on Automatic Control}, vol. 56, pp. 2639--2654, 2011.

\bibitem{Wang1}
Wang, L.Y. and Zhang, J.F. and Yin, G.G.
\newblock System identification using binary sensors.
\newblock {\em IEEE Trans. on Automatic Control}, vol. 48, pp. 1892--1907, 2003.

\bibitem{Wang2}
Wang, L.Y. and Yin, G.G. and Zhang, J.F.
\newblock Joint identification of plant rational models and noise distribution functions using binary-valued observations.
\newblock {\em Automatica}, vol. 42, pp. 543--547, 2006.

\bibitem{Koutsoukos}
Koutsoukos, X.D.
\newblock Estimation of hybrid systems using discrete sensors.
\newblock {\em Proceedings 42nd IEEE Conf. on Decision and Control}, pp. 155--160, 2003.

\bibitem{Aslam}
Aslam, J. and Butler, Z. and Constantin, F. and Crespi, V. and Cybenko, G. and Rus, D.
\newblock Tracking a moving object with a binary sensor network.
\newblock {\em Proceedings 1st ACM Conf. on Embedded Networked Sensor Systems}, pp. 150--161, 2003.

\bibitem{Djuric_2}
Djuric, P.M. and Vemula, M. and Bugallo, M.F.
\newblock Target tracking by particle filtering in binary sensor networks.
\newblock {\em IEEE Trans. on Signal Processing}, vol. 56, pp. 2229--2238, 2008.

\bibitem{Ristic}
Ristic, B. and Gunatilaka, A. and Gailis, R.
\newblock Achievable accuracy in Gaussian plume parameter estimation using a network of binary sensors.
\newblock {\em Information Fusion}, vol. 25, pp. 42--48, 2015.

\bibitem{Jazwinski}
Jazwinski, A.H.
\newblock Limited memory optimal filtering.
\newblock {\em IEEE Trans. on Automatic Control},  vol. 13, pp. 558--563, 1968.

\bibitem{RaoRawLee01}
Rao, C.V. and Rawlings, J.B. and Lee, J.H.
\newblock Constrained linear estimation--a moving horizon approach.
\newblock {\em Automatica},  vol. 37, no. 10, pp. 1619--1628, 2001.

\bibitem{RaRaMa03}
Rao, C.V. and Rawlings, J.B. and Mayne, D.Q.
\newblock Constrained state estimation for nonlinear discrete-time systems: stability and moving horizon approximations.
\newblock {\em IEEE Trans. on Automatic Control}, vol. 48, no. 2, pp. 246--257, 2003.

\bibitem{NLMHE}
Alessandri, A. and Baglietto, M. and Battistelli, G.
\newblock Moving horizon state estimation for nonlinear discrete-time systems: new stability results and approximation schemes.
\newblock {\em Automatica}, vol. 44, pp. 1753--1765, 2008.

\bibitem{AlBaBaZavCDC10}
Alessandri, A. and Baglietto, M. and Battistelli, G. and Zavala, V.M.
\newblock Advances in moving horizon estimation for nonlinear systems.
\newblock {\em Proceedings 49th IEEE Conference on Decision and Control}, pp. 5681--5688, 2010.

\bibitem{AlBaBaTAC05}
Alessandri, A. and Baglietto, M. and Battistelli, G.
\newblock Receding-horizon estimation for switching discrete-time linear systems.
\newblock {\em IEEE Trans. on Automatic Control}, vol. 50, no. 11, pp. 1736--1748, 2005.

\bibitem{GuoHuang13}
Guo, Y. and Huang, B.
\newblock Moving horizon estimation for switching nonlinear systems.
\newblock {\em Automatica}, vol. 49, no. 11, pp. 3270--3281, 2013.

\bibitem{FaFerrSca10}
Farina, M. and {Ferrari-Trecate}, G. and Scattolini, R.
\newblock Moving-horizon partition-based state estimation of large-scale systems.
\newblock {\em Automatica}, vol. 46, no. 5, pp. 910--918, 2010.

\bibitem{HabVerh13}
Haber, A. and Verhaegen, M.
\newblock Moving Horizon Estimation for Large-Scale Interconnected Systems.
\newblock {\em IEEE Trans. on Automatic Control}, vol. 58, no. 11, pp. 2834--2847, 2013.

\bibitem{SchnHannMarq15}
Schneider, R. and Hannemann-Tamás, R. and Marquardt, W.
\newblock An iterative partition-based moving horizon estimator with coupled inequality constraints.
\newblock {\em Automatica},  vol. 61, pp. 302--307, 2015.

\bibitem{Farina1}
Farina, M. and {Ferrari-Trecate}, G. and Scattolini, R.
\newblock Distributed moving horizon estimation for linear constrained systems.
\newblock {\em IEEE Trans. on Automatic Control},  vol. 55, no. 11, pp. 2462--2475, 2010.

\bibitem{Farina2}
Farina, M. and {Ferrari-Trecate}, G. and Scattolini, R.
\newblock Distributed moving horizon estimation for nonlinear constrained systems.
\newblock {\em International Journal of Robust and Nonlinear Control},  vol. 22, no. 2, pp. 123--143, 2012.

\bibitem{quantized_measurement}
Liu, A. and Yu, L. and Zhang, W.-A. and Chen, M.Z.Q.
\newblock Moving horizon estimation for networked systems with quantized measurements and packet dropouts.
\newblock {\em IEEE Trans. on Circuits and Systems I: Regular Papers},  vol. 60, pp. 1823--1834, 2013.

\bibitem{CDC15}
Battistelli, G. and Chisci, L. and Gherardini, S.
\newblock Moving horizon state estimation for discrete-time linear systems with binary sensors.
\newblock {\em Proceedings 54th IEEE Conf. Decision and Control}, pp. 2414--2419, 2015.

\bibitem{Wiener1}
Westwick, D. and Verhaegen, M.
\newblock Identifying {MIMO} {Wiener} systems using subspace model identification methods.
\newblock {\em Signal Processing}, vol. 52(2), pp. 235--258, 1996.

\bibitem{Wiener2}
Glaria L\'opez, T.A. and Sbarbaro, D.
\newblock Observer design for nonlinear processes with {Wiener} structure.
\newblock {\em Proceedings 50th IEEE Conf. Decision and Control and European Control Conference}, pp. 2211--2316, 2011.

\bibitem{BlMi}
Blanchini, F. and Miani, S.
\newblock Stabilization of {LPV} systems: State feedback, state estimation, and duality.
\newblock {\em SIAM Journal on Control and Optimization},  vol. 42, pp. 76--97, 2003.

\bibitem{PWCP}
Louveaux, F.V.
\newblock Piecewise convex programs.
\newblock {\em Mathematical Programming}, vol. 15(1), pp. 53--62, 1978.

\bibitem{OPT1}
Lau, K.K. and Womersley, R.S.
\newblock Multistage quadratic stochastic programming.
\newblock {\em Journal of Computational and Applied Mathematics}, vol. 129(1-2), pp. 105--138, 2001.

\bibitem{OPT2}
Patrinos, P., and Sarimveis, H.
\newblock Convex parametric piecewise quadratic optimization: Theory and algorithms.
\newblock {\em Automatica}, vol. 47(8), pp. 1770--1777, 2011.

\bibitem{bounds}
Wu, L.
\newblock Error bounds for piecewise convex quadratic programs and applications.
\newblock {\em SIAM Journal on Control and Optimization}, vol. 33(5), pp. 1510--1529, 1995.

\end{thebibliography}
\end{document}